\documentclass[10pt,twocolumn,floatfix,prl,superscriptaddress]{revtex4-1}
\usepackage{graphicx}
\usepackage{color}
\usepackage{xcolor}
\usepackage{amssymb}
\usepackage{amsmath}
\usepackage{epsfig}
\usepackage{bm}
\usepackage[american]{babel}
\usepackage{hyperref}
\usepackage{braket}
\usepackage[T1]{fontenc} 
\usepackage{txfonts}
\usepackage{comment}
\usepackage{textcomp}

\DeclareGraphicsExtensions{.pdf,.png,.jpg}
\hypersetup{colorlinks=true,urlcolor=blue,breaklinks=true,citecolor=blue,linkcolor=blue,pdfstartview=FitH,pdfpagemode=UseNone}

\usepackage{physics}

\addto\captionsenglish{}

\begin{document}

\title{High-fidelity multi-photon-entangled cluster state with solid-state quantum emitters in photonic nanostructures}

\date{\today}

\author{Konstantin Tiurev} \email{konstantin.tiurev@gmail.com}
\affiliation{Center for Hybrid Quantum Networks (Hy-Q), The Niels Bohr Institute, University~of~Copenhagen,  DK-2100  Copenhagen~{\O}, Denmark}

\author{Martin~Hayhurst~Appel}
\affiliation{Center for Hybrid Quantum Networks (Hy-Q), The Niels Bohr Institute, University~of~Copenhagen,  DK-2100  Copenhagen~{\O}, Denmark}

\author{Pol~Llopart~Mirambell}
\affiliation{Center for Hybrid Quantum Networks (Hy-Q), The Niels Bohr Institute, University~of~Copenhagen,  DK-2100  Copenhagen~{\O}, Denmark}

\author{Mikkel~Bloch~Lauritzen}
\affiliation{Center for Hybrid Quantum Networks (Hy-Q), The Niels Bohr Institute, University~of~Copenhagen,  DK-2100  Copenhagen~{\O}, Denmark}

\author{Alexey~Tiranov}
\affiliation{Center for Hybrid Quantum Networks (Hy-Q), The Niels Bohr Institute, University~of~Copenhagen,  DK-2100  Copenhagen~{\O}, Denmark}

\author{Peter~Lodahl}
\affiliation{Center for Hybrid Quantum Networks (Hy-Q), The Niels Bohr Institute, University~of~Copenhagen,  DK-2100  Copenhagen~{\O}, Denmark}

\author{Anders~S{\o}ndberg~S{\o}rensen}
\affiliation{Center for Hybrid Quantum Networks (Hy-Q), The Niels Bohr Institute, University~of~Copenhagen,  DK-2100  Copenhagen~{\O}, Denmark}

\begin{abstract}
We propose a complete architecture for deterministic generation of entangled multiphoton states. Our approach utilizes periodic driving of a quantum-dot emitter and an efficient light-matter interface enabled by a photonic crystal waveguide. We assess the quality of the photonic states produced from a real system by including all intrinsic experimental imperfections. Importantly, the protocol is robust against the nuclear spin bath dynamics due to a naturally built-in refocussing method reminiscent to spin echo. We demonstrate the feasibility of producing  Greenberger--Horne--Zeilinger and one-dimensional cluster states with fidelities and generation rates exceeding those achieved with conventional `fusion' methods in current state-of-the-art experiments. The proposed hardware constitutes a scalable and resource-efficient approach towards implementation of measurement-based quantum communication and computing.
\end{abstract}

\maketitle
The development of efficient sources of on-demand entangled photons is an ongoing experimental endeavour. Quantum states containing large numbers of entangled photons is a desirable component for many quantum-information processing applications, including photonic quantum computing~\cite{RevModPhys.79.135,PhysRevLett.93.040503,PhysRevLett.95.010501,Knill:2001aa,doi:10.1063/1.5115814,doi:10.1063/1.4976737}, quantum simulations~\cite{Lanyon:2010aa,Ma:2011aa}, entanglement-enhanced metrology~\cite{T_th_2014,shettell2019graph}, and long-distance quantum communication~\cite{Azuma:2015aa,Li:2019aa,Buterakos2017,hilaire2020resource,PhysRevX.10.021071,PhysRevA.95.012304}. Furthermore, access to high-fidelity multiphoton entanglement would have applications for fundamental tests of quantum mechanics~\cite{Pan:2000aa,Lu.2014,PhysRevA.61.022109}.

The creation of entangled states containing large numbers of photons is, however, a formidable challenge due to the lack of deterministic and scalable methods for the production of such states. Spontaneous parametric downconversion~(SPDC) sources~\cite{PhysRevLett.25.84,PhysRevLett.75.4337,PhysRevLett.83.3103} combined with interference between generated pairs and single photon detection~\cite{PhysRevLett.82.1345,PhysRevA.73.022330,PhysRevLett.78.3031} have been implemented to scale up the number of entangled photons~\cite{PhysRevLett.95.010501,PhysRevLett.95.210502,Zhang:2019aa,Lu:2007aa,Yao:2012aa,PhysRevLett.117.210502}, with a recent state-of-the art experiment demonstrating genuine 12-photon entanglement~\cite{PhysRevLett.121.250505}. Today, scaling up is pursued also commercially by multiplexing many probabilistic SPDC sources towards photonic quantum computing~\cite{doi:10.1063/1.4976737,doi:10.1063/1.5115814}. An alternative and much less investigated strategy is to apply on-demand photon emission from a single quantum emitter. In this case, a single spin in the emitter serves as the entangler of consecutively emitted photons~\cite{PhysRevA.58.R2627,Lindner2009,Lee2019,Buterakos2017,hilaire2020resource,PhysRevX.10.021071,Segovia2019,Economou2010}, and combined with photonic nanostructures for enhancing photon-emitter coupling~\cite{Lodahl2015}, long strings of highly-entangled photons could potentially be generated. A proof-of-concept experiment with quantum dots (QDs) in bulk samples recently demonstrated three-qubit linear cluster states~\cite{Schwartz434}. However, it is an open question how these deterministic sources can be scaled-up in a real experimental setting. 
\begin{figure*}[t]	
\includegraphics[width=0.9\textwidth]{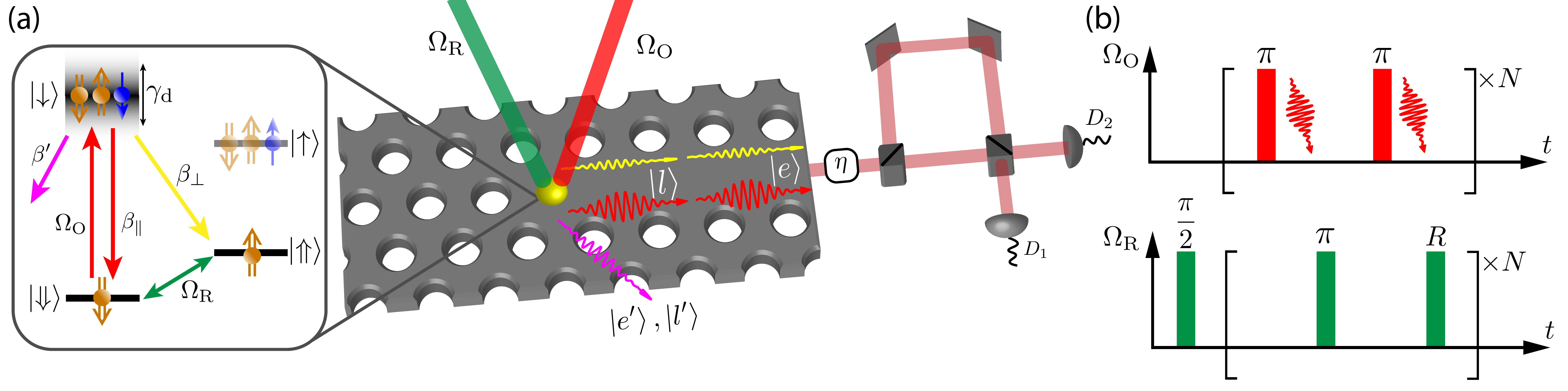}
\caption{\label{fig:1} {Architecture} for generation and measurement of time-bin entangled photons. (a)~Center:~Light-matter interface based on a QD~(yellow dot) placed in a PCW formed of semiconductor with a periodic dielectric structure. Driven cyclically by excitation pulses $\Omega_{\mathrm{O}}$~(red beam) and rotation pulses $\Omega_{\mathrm{R}}$~(green beam), a QD emits entangled photonic qubits in either the early or the late temporal mode. Inset:~Energy level structure of a positively charged QD consisting of hole spin states $\ket{\Uparrow}$ and $\ket{\Downarrow}$ and trion states $\ket{\uparrow}$ and $\ket{\downarrow}$. Upon spontaneous emission, an early~($\ket{e}$) or a late~($\ket{l}$) photon is emitted into the PCW on either the vertical transition~(red decay path in inset/red wave in the PCW), or on the undesired diagonal transition~(yellow decay path/yellow wave). Additionally, an early~($\ket{e^{\prime}}$) or a late~($\ket{l^{\prime}}$) photon can be emitted out of the waveguide mode and thereby lost~(purple decay path/purple wave). The PCW simultaneously ensures a high internal efficiency $\beta_{\parallel} + \beta_{\perp}$ as well as a high selectivity of the vertical decay path. Right: Setup for detection of time-bin entangled photons. Passing photons through a single interferometer arm yields a Z-measurement while interfering the early and late photons at the final beam splitter (either via passive or active routing) yields a measurement in the X or Y basis. Here $\eta$ represents the total measurement efficiency. 
(b)~Sequence of pulses $\Omega_{\mathrm{O}}$ and $\Omega_{\mathrm{R}}$ used to generate time-bin-encoded entangled photonic states. Optical pulses $\Omega_{\mathrm{O}}$ are used to excite a transition $\ket{\Downarrow} \rightarrow \ket{\downarrow}$ with radiative decay rate $\gamma$, while the ground-state rotations $\Omega_{\mathrm{R}}$ are realized with Raman transitions.}
\end{figure*}

In the present Letter we develop a complete architecture for deterministic generation of time-bin entangled multi-photon states. Our proposal exploits a QD emitter embedded in a photonic nanostructure and removes the dominant noise source through a built-in spin-echo protocol. We investigate the performance of the architecture, taking into account all intrinsic imperfections and identify the governing physical processes and figures-of-merit, hence providing a path for scaling-up the protocol. Our results demonstrate that recent experimental advances make QDs in photonic nanostructures highly promising sources of scalable multiphoton entangled states.

Self-assembled semiconductor QDs have lately seen remarkable experimental progress, opening new possibilities for photonic quantum technologies. Particularly, spin qubits realized with a single charge injected into the QD enable efficient coherent light-matter interfaces and control over emitted photons due to simultaneously achievable high photon generation rate, good optical and spin coherence properties~\cite{Aharonovich2016,Atature2018,Awschalom2018,Lodahl2015}, and near-perfect spin-rotations~\cite{Bodey:2019aa}. 
Integration of QDs into photonic nanostructures, such as photonic crystal waveguides~(PCW), significantly improves the quality of quantum interfaces and has resulted in single-photon indistinguishability~($I$) of two subsequently emitted photons exceeding 96\%~\cite{Kirsanske2017,Lodahl2015,Trotta2017,PhysRevB.100.155420,PhysRevLett.116.020401,tomm2020bright}, an internal efficiency $\beta$ exceeding 98\%~\cite{Arcari2014} and on-demand entangled photon sources with higher than 90\% state fidelity~\cite{Wang2019}. Recently it was demonstrated that these sources can be scaled up to reach the threshold for quantum advantage~\cite{Uppueabc8268}. 

The proposed architecture based on a QD containing a hole spin in a PCW is illustrated in Fig.~\ref{fig:1}. 
It relies on encoding photonic qubits in separate time bins corresponding to early $(\left| e \right>)$ or late $(\left| l \right>)$ arrival times. The general idea is to repeatedly apply the pulse sequence of Fig.~\ref{fig:1}(b) to coherently control a ground-state spin in the QD and emit single photons on the targeted optical transition in the designated time bin. Initially the hole spin is prepared in a superposition of the two spin states $\ket{\Downarrow}$ and $\ket{\Uparrow}$ using a $\pi/2$ pulse from the Raman field $\Omega_{\mathrm{R}}$. Within each round of the protocol the QD is first excited to the trion state $\ket{\downarrow}$ using the optical field $\Omega_{\mathrm{O}}$ if the QD is in $\ket{\Downarrow}$. From the trion state the QD decays emitting an early photon $\ket{e}$. Subsequently the hole spins states are flipped using a Raman $\pi$-pulse followed by excitation with $\Omega_{\mathrm{O}}$ and emission of a late photon $\ket{l}$. This procedure creates an entangled state between the spin and the time bin of the outgoing photon, which can be extended to multiple photons by repeating the protocol with a spin rotation $R$ between each round of the protocol. The nature of the entangled  state  is defined by the  choice of $R$: $R = \pi$ creates the Greenberger--Horne--Zeilinger~(GHZ) state~\cite{greenberger2007going} while $R = \pi/2$ creates the one-dimensional cluster state~\cite{Tiurev2019a}.

A similar scheme has been considered experimentally in Ref.~\cite{Lee2019} using a QD in a micropillar cavity. That work, however, was unable to generate entanglement due to the inability to drive spin rotations while having cycling optical transition. Our proposal solves this problem by embedding a quantum dot into a photonic crystal waveguide. As shown in a recent experiment~\cite{PhysRevLett.126.013602}, high-quality optical cyclings can be induced on the designated transition due to the high coupling asymmetry of the two in-plane linear dipole transitions. In combination with ground-state spin rotations, this is the basis for the efficient scaling of the protocol. In the following, we prove and benchmark the scalability of the approach by evaluating the fidelity of multi-photon GHZ and cluster states in the presence of all relevant imperfections.

\begin{figure*}[t]	
\includegraphics[width=0.95\textwidth]{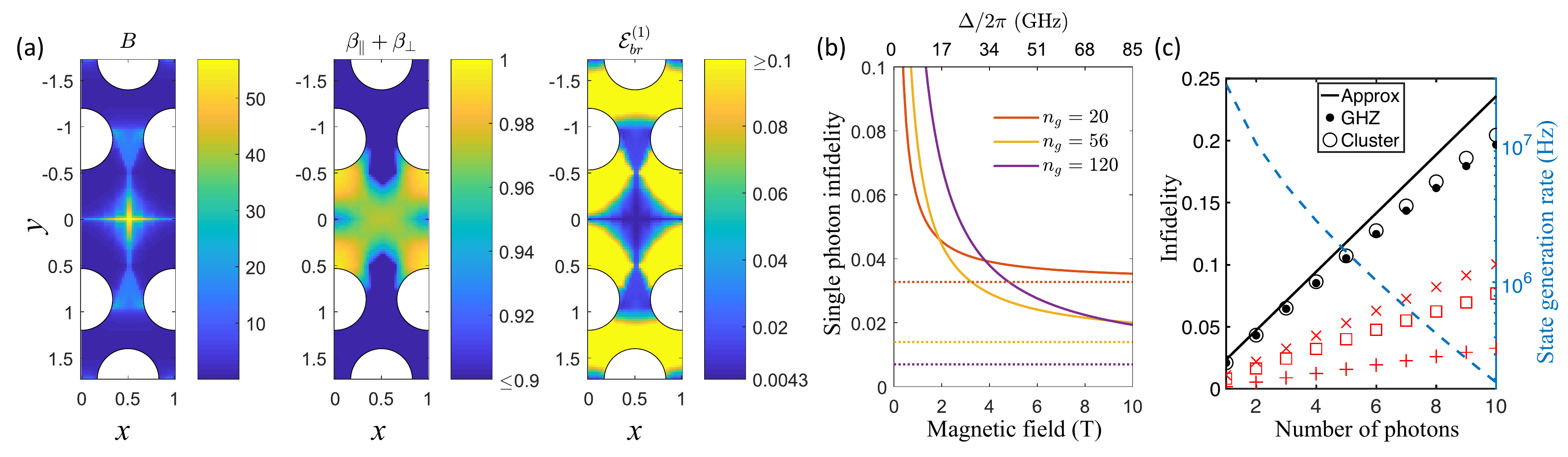}
\caption{\label{fig:2} (a)~Spatial map of the optical branching ratio asymmetry $B$~(left), total coupling efficiency into the mode of the waveguide $\beta_\parallel + \beta_\perp$~(center), and the infidelity of the spin-photon entangled state due to the spatially varying branching ratio~(right) within the unit cell of the PCW for $n_g=20$. (b)~Infidelity of spin-photon entangled states versus the applied magnetic field~(detuning $\Delta$ corresponding to a $g$-factor of $|g| = |g_e| + |g_h| = 0.6$~\cite{PhysRevB.91.165304}) for various group indices $n_g$ attributed to different spectral positions with respect to the band gap of the PCW. Dashed lines mark the infidelities in the limit $\Delta \rightarrow \infty$. For each $n_g$, $\gamma$ is evaluated using the simulated Purcell factor assuming a bulk rate of $\gamma_{\mathrm{bulk}}=1$~ns$^{-1}$~\cite{Lodahl2015}. (c)~Total infidelity versus number of photons with all imperfections taken into account and $n_g = 56$. The solid line shows the infidelity in the first-order approximation~\eqref{eq:infidelity_total}, while the black symbols are beyond perturbation theory~\cite{Tiurev2019a}. Red symbols show the different contributions in Eq.~\eqref{eq:infidelity_total}, i.e. dephasing $\mathcal{E}_{\mathrm{ph}}$~(\textcolor{red}{$\times$}), excitation errors ${\mathcal{E}}_{\mathrm{exc}}$~(\textcolor{red}{$\Box$}), and imperfect branching ${\mathcal{E}}_{\mathrm{br}}$~(\textcolor{red}{$+$}). The dashed line shows the state generation rate for an outcoupling efficiency $\eta = 0.84$~\cite{Uppueabc8268} and a cycle length $T_{\mathrm{cycle}} = 27$~ns. See main text for parameters.} 
\end{figure*}

We assess the quality of the spin-multiphoton state by calculating the infidelity~\cite{Tiurev2019a} $\mathcal{E}^{(N)} = 1 - \mathrm{Tr}_{\mathrm{env}} \{ \bra{\Psi} \hat{\rho}^{(N)} \ket{\Psi} \}$, where $\hat{\rho}^{(N)}$ is the density operator of an $N$-photon state affected by imperfections, $\ket{\Psi}$ is the target state, and $\mathrm{Tr}_{\mathrm{env}}$ denotes a trace over the emission time and unobserved degrees of freedom, such as phonons or lost photons. Photonic quantum information protocols, e.g,  for quantum communication~\cite{Azuma:2015aa,Li:2019aa,Buterakos2017,hilaire2020resource,PhysRevX.10.021071} or computation~\cite{PhysRevLett.115.020502,PhysRevLett.100.060502} are typically designed to be loss-tolerant by postselecting events without photon loss. In this case the relevant quantity is the fidelity conditioned on the detection of at least one photon in either the early or late time bin. The conditional infidelity for the generation of an entangled GHZ or cluster state containing N photons and the spin is in first-order perturbation theory given by~\cite{Tiurev2019a}
\begin{equation}
\begin{aligned}
\label{eq:infidelity_total}
    \mathcal{E}^{(N)}
    &=
    N\Big{(}
    \frac{1 - I}{2}
    +
    \frac{\sqrt{3}\pi}{8}\frac{\gamma}{\Delta}
    +
    \frac{1}{2(B+1)}
    \Big{)}
    -
    \frac{1}{4(B+1)}.
\end{aligned}
\end{equation}
Here the spontaneous emission rate $\gamma$, the branching ratio $B$, the degree of indistinguishability $I$, and detuning of the off-resonant transition $\Delta$ are parameters that will be explained below.

The ideal protocol assumes that only the vertical decay path $\ket{\downarrow} \rightarrow \ket{\Downarrow}$ in Fig.~\ref{fig:1}(a) 
is allowed, such that the excitation and decay form a closed cycle. A finite probability of the diagonal transition $\ket{\downarrow} \rightarrow \ket{\Uparrow}$ will lead to an incorrect spin configuration and a reduction of the fidelity. 
We characterize the cyclicity with a branching parameter $B = (\beta_{\parallel} + \beta_{\parallel}^{\prime})/(\beta_{\perp} + \beta_{\perp}^{\prime})$, where $\beta_{\parallel}$($\beta_{\perp}$) and $\beta_{\parallel}^{\prime}$($\beta_{\perp}^{\prime}$) are the probabilities of the vertical(diagonal) transitions into and out of the waveguide mode, respectively. The performance of an experiment will therefore rely on the high selectivity of the vertical transitions, i.e. $B\gg 1$. 

Using an out-of-plane magnetic field (Faraday configuration) would almost fully eliminate the diagonal transitions due to the selection rules and therefore allow for near-perfect branching conditions, but would at the same time prohibit the coherent ground-state Raman transition $\Omega_{\mathrm{R}}$, which the protocol depends upon. Instead, we consider an in-plane magnetic field (Voigt geometry) which provides $B=1$ in bulk but, crucially, allows all-optical spin control. $B$ may then be increased by selectively enhancing the desired optical transition with a photonic nanostructure~\cite{PhysRevLett.126.013602,Carter2013,Sun2016,Lee2019,Wang2019a}. 
This, in combination with the orthogonally polarised linear dipoles of a QD in the Voigt geometry, allows a greatly enhanced $B$ while simultaneously retaining the possibility of ground-state spin rotations. In Fig.~\ref{fig:2}(a) we show calculated $\beta$-factors and branching ratios $B$ based on the field distribution calculated in Ref.~\cite{Javadi:18}. For a realistic group index $n_g=20$, which specifies the PCW-induced slow-down factor, a branching ratio of $B>50$ and an internal efficiency $\beta > 96\%$ are simultaneously achievable by placing a QD in the center of a PCW. To further suppress the residual contribution of the diagonal transitions, we consider frequency filters which can be implemented without introducing significant loss~\cite{Uppueabc8268} using e.g. narrow bandpass external (i.e. off-chip) etalons. Assuming high-efficiency filtering of the off-resonant photons, we derive~\cite{Tiurev2019a} the first-order infidelity due to imperfect branching ${\mathcal{E}}^{(N)}_{\mathrm{br}} = (N-1/2)/(2(B+1))$, which corresponds to the last two terms in Eq.~\eqref{eq:infidelity_total} and is shown in Fig.~\ref{fig:2}(a) for a single emitted photon. For the optimal QD position the single-photon branching infidelity can be lower than $1\%$.

Next, we consider the effect of dephasing omnipresent in solid state systems. Decoherence appears through a  variety of mechanisms characterized by widely different time scales. As discussed below the protocol is remarkably insensitive to slowly varying processes. On the other hand, phonon scattering appears on timescales~($\sim$ ps) shorter than the lifetime of the excited trion state~($\sim$ ns) and determines the indistinguishability of emitted photons~\cite{PhysRevLett.120.257401,PhysRevB.63.155307,PhysRevLett.93.237401,PhysRevLett.116.033601}. 
Pure dephasing broadens the zero-phonon line, resulting in the indistinguishability of emitted photons $I = \gamma/(\gamma+2\gamma_{\mathrm{d}})$, where $\gamma_{\mathrm{d}}$ is the phonon-induced dephasing rate~\cite{PhysRevLett.120.257401}. The same mechanism increases the state infidelity, which can hence be expressed through the experimentally measurable indistinguishability $I$ as $\mathcal{E}_{\mathrm{ph}}^{(N)} = N(1 - I)/2$, corresponding to the first term in Eq.~\eqref{eq:infidelity_total}.

Further, we discuss imperfect  operations during the driving pulses. Since the excited trion comprises two Zeeman states~[Fig.~\ref{fig:1}(a)], excitation of undesired transitions have to be suppressed. This is ensured by a large detuning $\Delta$ of the off-resonant transition $\ket{\uparrow} \leftrightarrow \ket{\Uparrow}$ compared to the decay rate $\gamma$ of the $\ket{\downarrow} \leftrightarrow \ket{\Downarrow}$ transition. The detuning can be controlled by a magnetic field, while $\gamma$ can be controlled via the Purcell effect of the waveguide. The probability of off-resonant excitations is strongly suppressed when the system is driven with long and low-intensity laser pulses. 
On the other hand for long pulses there is a large probability for the desired $\ket{\downarrow} \leftrightarrow \ket{\Downarrow}$ transition to decay and be re-excited during the pulse. The duration of the pulse should thus be optimized to suppress the errors. 
We have evaluated~\cite{Tiurev2019a} the infidelities corresponding to the optimal driving regimes for both Gaussian and square-shape pulses. The latter allows for a simple analytical expression, $\mathcal{E}^{(N)}_{\mathrm{exc}}=N\sqrt{3}\pi\gamma/(8\Delta)$, which also represents a good approximation for Gaussian pulses. Additional errors occur if the excitation laser drives the cross transitions $\ket{\Uparrow} \leftrightarrow \ket{\downarrow}$ and $\ket{\Downarrow} \leftrightarrow \ket{\uparrow}$, which, however, can be completely avoided by correct laser polarisation in side channel excitation. This is readily implementable in the waveguide geometry~\cite{Uppueabc8268} but has not yet been implemented in micropillar~\cite{Wang2019a} or planar cavities~\cite{Carter2013,Sun2016}, which rely on cross excitation schemes. 

Finally, we consider dephasing induced by slow drifts of the energy levels. A particular example arises from the hyperfine interaction between the coherent spin and the slowly fluctuating nuclear spin environment, i.e. the Overhauser noise~\cite{PhysRevLett.95.076805,PhysRevLett.88.186802}, which manifests itself in relatively short ground-state spin coherence times $T_2^*$~\cite{HU:2002aa}. Strikingly, our protocol for time-bin photon generation is insensitive to dephasing induced by such mechanism, because the pulse sequence of Fig.~\ref{fig:1}(b) flips the ground states $\ket{\Uparrow}$ and $\ket{\Downarrow}$ between the early and late photons. We assume that the photons are subsequently analyzed using the measurement setup of Fig.~\ref{fig:1}(a), which interferes pulses delayed by a time equal to the time difference between the two excitation pulses. Since the interferred components have spent exactly the same amount of time in the excited states for the early and late parts, our protocol inherently implements a perfect spin echo sequence~\cite{PhysRevLett.100.236802} at each cycle of the protocol without need for any additional refocusing methods. The slow drift of the central frequency of the transition will hence not have any influence on the interference. Consequently, either hole or electron spins can be used on an equal basis, even though the latter has a much shorter coherence time $T_2^*$. On longer times, slow fluctuations of the environment build up to a so-called $T_2$ noise. This, however, typically happens on time scales~\cite{PhysRevB.97.241413,Press2010} two orders of magnitude longer than the length of a cycle $T_{\mathrm{cycle}}$~\cite{Jayakumar:2014aa}, adding an error $\propto (T_{\mathrm{cycle}}/T_2)^2$ negligible compared to other imperfections.

The insensitivity to slow fluctuations is linked to the measurement setup in Fig.~\ref{fig:1}(a) where pulses from a single QD are interfered, but does not apply if  e.g.  attempting to fuse cluster states emitted by different QDs~\cite{PhysRevLett.82.1345,PhysRevA.73.022330,PhysRevLett.78.3031}. Nevertheless it captures several interesting situations. The quantum repeater protocol of Borregaard~\textit{et al.}~\cite{PhysRevX.10.021071} exploits a single emitter to produce entangled states containing hundreds of photons. Of these, only one photon is interfered with a different emitter, while the remaining $N-1$ photons are measured using the setup in Fig.~\ref{fig:1}(a) and hence fulfill the effective spin echo conditions. Similarly, the protocol of Pichler~\textit{et al.}~\cite{Pichler11362} for universal quantum computation using cluster states relies on the emission from a single emitter. We thus expect a similar robustness.

All error terms in Eq.~\eqref{eq:infidelity_total} depend on the group index of the waveguide: a high $n_g$ increases the decay rate $\gamma$ and hence the indistinguishability, but at the same time results in stronger driving of off-resonant transition. Furthermore, the branching ratio can also be improved by the enhancement of $n_g$. The waveguide therefore can be used to control the trade-off between errors and optimize the output state. As shown in Fig.~\ref{fig:2}(b), a high $n_g$ becomes beneficial given sufficient Zeeman splitting, i.e. for a strong magnetic field or large $g$-factor~\cite{doi:10.1063/1.3367707,PhysRevLett.112.107401,PhysRevB.91.165304}. By engineering the photonic crystal band gap and increasing the group indices to higher values, the single spin-photon infidelity can be reduced to the levels of $\approx$0.5\% for sufficiently strong magnetic fields, as shown with dashed lines in Fig.~\ref{fig:2}(b). For more modest magnetic fields, a spin-photon entangled state fidelity above $95\%$ can be reached.

The case of $N=3$ is of special importance since it potentially serves as a building block for photonic quantum {protocols}~\cite{PhysRevA.95.012304,PhysRevLett.115.020502,doi:10.1063/1.4976737}. Such three-photon states can also be realized by fusing six single photons with a total probability of $1/32$~\cite{PhysRevLett.100.060502}. With state-of-the-art SPDC 
sources operating at MHz frequencies and an extraction efficiency of $\approx$70\%~\cite{Kanedaeaaw8586}, the theoretical three-photon GHZ state generation rate is in the few kHz regime. In comparison, using a deterministic source with the parameters of Ref.~\cite{Uppueabc8268} we estimate a direct three-photon production rate of $\approx$5~MHz~[see Fig.~\ref{fig:2}(c)], which exceeds the estimate for SPDC-based method by three orders of magnitude. The fidelity of such three-photon states is $\approx$83\% for experimentally measured parameters, $B=15$~\cite{PhysRevLett.126.013602}, $\Delta = 2\pi\times 16$~GHz~corresponding to a magnetic field of 2 T~\cite{PhysRevLett.126.013602,Prechtel:2015aa}, $\gamma = 3.2$~ns$^{-1}$, and single-photon indistinguishability $I = 0.96$~\cite{PhysRevLett.126.013602} corresponding to $\gamma_{\mathrm{d}} = 0.06$~ns$^{-1}$~\cite{Uppueabc8268}. Figure~\ref{fig:2}(c) also shows the infidelity corresponding to improved experimental parameters, $B=140$ calculated from the field distributions of Ref.~\cite{Javadi:18} for $n_g = 56$, $\Delta = 2\pi\times 64$~GHz, $\gamma_{\mathrm{d}} = 0.06$~ns$^{-1}$~\cite{Uppueabc8268}, and $\gamma = 5.3$~ns$^{-1}$ ($I = 0.98$~\cite{PhysRevLett.126.013602,Lodahl2015}). 

For large entangled states the infidelity inherently grows with the system size. Hence the relevant quantity for quantum-information applications is the infidelity per qubit. In a gate based approach fault tolerant thresholds of  3.2\% (1.4\%) single qubit error has been derived for  quantum computation with 3D (2D) cluster states~\cite{RAUSSENDORF20062242,PhysRevLett.98.190504,Raussendorf_2007}. Potentially our deterministic generation scheme can be mapped to these setups using gates between QDs~\cite{Economou2010} or efficient Bell state measurements of photons~\cite{Witthaut_2012}. 
%
For the  parameters of Fig.~\ref{fig:2}(c), we find a comparable  infidelity of 2.1\% per qubit, consisting of 1.8\% and 0.3\% of single- and 2-qubit errors, respectively. A full assessment of this possibility should take into account additional errors  when extending cluster states to higher dimensions and the need for post-selection but also advantages from knowing the error mechanism and the ability to do long range interactions. This is beyond the scope of this article, but it is encouraging that the numbers we obtain for realistic experimental parameters are comparable to these requirements. The states can also be extended to higher dimensions using linear optics based fusion, but in this case the error thresholds are more stringent~\cite{bartolucci2021fusionbased}.  Alternatively, the requirements for  quantum communication protocols~\cite{PhysRevA.98.052320,ASIACRYPT2005,ASIACRYPT2007,PhysRevA.59.1829,7935fdc5be87497bb613fe4bbb79a6ad}
are typically much more relaxed. 
Taking the security analysis of the anonymous transmission protocol~\cite{PhysRevA.98.052320} as an example, the predicted error rates are within the threshold for up to at least fifty parties and almost an order of magnitude below the threshold for four parties.

In conclusion, we have proposed a complete architecture of a device for scalable multiphoton entanglement generation from a QD-based emitter. Our particular implementation relies on the control of photon emission by means of nanophotonic structure, such as PCWs. Our findings predict near-future feasibility of multiphoton sources with encouraging state fidelities and generation rates compared with existing methods. The provided theoretical analysis improves our understanding of the mechanisms governing the quality of the produced states and provides a recipe for further improvement of such devices.

\begin{acknowledgments}
We gratefully acknowledge financial support from Danmarks Grundforskningsfond~(DNRF 139, Hy-Q Center for Hybrid Quantum Networks), the European Research Council (ERC Advanced Grant `SCALE'), and the European Union Horizon 2020 research and innovation programme under grant agreement N\textsuperscript{\underline{o}}~820445 and project name Quantum Internet Alliance.
\end{acknowledgments}

\bibliographystyle{apsrev4-1}
\bibliography{reflist}
\end{document}